\documentclass{article}

\usepackage{arxiv}

\usepackage[utf8]{inputenc} 
\usepackage[T1]{fontenc}    
\usepackage{hyperref}       
\usepackage{url}            
\usepackage{booktabs}       
\usepackage{amsfonts}       
\usepackage{nicefrac}       
\usepackage{microtype}      
\usepackage{lipsum}		
\usepackage{graphicx}
\usepackage{natbib}
\usepackage{doi}

\title{Fast, Cheap, Good:\\ Lightweight Methods Are Undervalued}

\author{
 Adam Shostack\\
 Shostack + Associates, Inc and
  University of Washington\\
  Seattle, Washington 98195 \\
  \texttt{adam@shostack.org} \\
}

\renewcommand{\headeright}{(Preprint)}
\renewcommand{\undertitle}{(Preprint)}
\renewcommand{\shorttitle}{Fast, Cheap, Good: Lightweight Methods are Undervalued}

\hypersetup{
pdftitle={Fast, Cheap, Good: Lightweight Methods are Undervalued},
pdfsubject={q-bio.NC, q-bio.QM},
pdfauthor={Adam Shostack},
pdfkeywords={security, thread modeling, security engineering, usable security},
}

\begin{document}
\maketitle

\begin{abstract}
Engineering techniques to address the endless parade of security issues are an
important area of research. Properties of practices in industrial use are rarely studied.
Security workers satisfice. There is a widespread perception that security work must be cumbersome, and thus there’s no value to assessing levels of effort. This is complemented by a belief that the nth day of work will produce value equal to the first. These perceptions impact both practice and research. This paper expands the acceptable paradigms for security analysis to include the fast, cheap and good enough. “Nothing” is often enough for industry. This paper makes a case for valuing lightweight (“fast and cheap”) methods, presents a set of case studies and evaluation criteria for such tools, including card decks and role playing games.
\end{abstract}


\section{Introduction}
“Copying and Pasting code from Stack Exchange” is a derisive meme, built on the idea that there’s a Proper Solution, and ... there’s what you find on Stack Exchange. This behavior – looking for quick solutions – demonstrates two realities. The first is even skilled technical professionals find themselves needing help in unfamiliar situations. Second, efficiency is
often scorned.

This paper explores a new frame for security methodologies: fast and cheap are good.\footnote{Forthcoming as \emph{Nothing is Good Enough} IEEE S+P Magazine, 2023} This
is in intentionally provocative contrast to the engineering truism of “fast, cheap, good: choose any two.” We are not arguing these tradeoffs don’t exist, but the aphorism exists because demands for such tradeoffs are common. Research into fast and cheap tools is undervalued relative to mathematically or otherwise “sophisticated” tools. Approaches that prioritize speed or simplicity are frequently derided as trivial or obvious, rather than praised as elegant, or recognized as what security workers need.

The new paradigm leads us to ask a new question about the distribution of security issues (vulnerabilities or flaws). If such issues are very similar to one another, then knowledge of the system is important, because the primary effort to find the issues is search – where might this crop up? If such issues are unique, then security knowledge or even ‘an adversarial mindset’ may be paramount.

I and others create low-friction threat modeling methodologies. This work grounds and investigates the intuition that lower friction is a worthwhile goal for security workers. This paper explains threat modeling and tools in response to constraints. We analyze three tools we label fast, cheap and good (Section 3). From there, we construct a framework, a flaw space and an Flaw Oracle which reliably finds them (Section 4). The juxtaposition enables us to ask how we should judge them. We propose initial criteria of cost, expertise requirements and quality (Section 5).

\subsection{Contribution}
This work:

\begin{enumerate}
    \item Collects and organizes a set of new techniques that are rarely discussed coherently
in the literature.
    \item Produces a framework, motivated by those new techniques, that explicitly considers
the difficulty of threat modeling practice.
    \item Provide evaluation criteria for new techniques this framework allows.
    \item Ask if vulnerabilities are more like products of a factory, very similar to each other, or artisanal, each unique and special.
\end{enumerate}

\section{Background}
People satisfice. They accept “good enough” except where there are unavoidable pressures
to go further. By way of example, “we do not use any of our limited citation count for the idea of satisficing.”

Threat modeling is a family of techniques to anticipate future problems with a system, so
they can be addressed. There are many competing methodologies and evaluation criteria vary widely. In practice, threat modeling students often express concerns that it will fail to find all relevant threats or complete in appropriate time. 
\footnote{Personal observation as an instructor}

Is one possible explanation—an explanation that could implicitly undergird all of these
observations—that threat modeling is or appears harder than it needs to be? Perhaps
threat modeling practices, designed to be comprehensive, become so comprehensive that they are too tedious to learn or execute? (This is a ``perfect is the enemy of the good'' theory.)

What happens when a threat modeling practice trades precision for ease, becoming less
burdensome, but easier to perform and more accessible to a wide range of practitioners?
Prior work, which we discuss in section 3, suggests that the answer can be that practices become more effective. Can these practices, or ones similar to them, actually improve threat modeling overall, leveraging their relative light `weight' to help practitioners perform threat modeling more regularly than they would otherwise? Through our review of this literature, we argue that the answer to this question is also yes. Building on this argument, the next section builds a conceptual framework for “fast, cheap, and good enough” threat
modeling practices.


\subsection{Related work}
There is a rich literature on the use of serious games or games with a purpose outside the scope of this paper. For example, (\cite{Prensky2006}) and (\cite{Abt1987}).
There now is a well-understood set of challenges for usable security, including that security is often a secondary task. For this preprint, we assume that software engineering and systems operations are security workers, at least some of the time. We assume that those engineers have a compliance budget (\cite{beautement_sasse_wonham_2008}), and that faster and cheaper techniques are more likely to fit within it. We assume that for those workers, the larger the request, the more resistance it will meet, and conversely, it is easier to get them to do easier work. That is: fast and cheap are good because security workers satisfice. (\cite{Garfinkel2014})

\subsection{Software is used in diverse ways}

There are many dimensions of diversity of software:
\begin{enumerate}
    \item Development lifecycles range from continuous development and deployment
    through agile sprints to waterfall.
    \item Programming languages range from assembler to Javascript or even codeless
    systems.
    \item Deployment environments range from cloud systems through downloaded software through IoT and even interplanetary spacecraft.
    \item Impact of failure ranges from the very low impact of a local game failing through medical device software failure potentially costing many lives.
\end{enumerate}

Perhaps threat modeling practices, designed to address the highest risk systems, become so comprehensive that they are too tedious to learn or execute?  What happens when a threat modeling practice trades precision for ease, becoming less burdensome, but easier to perform and more accessible to a wide range of practitioners? Prior work, which we discuss in section 3, suggests that the answer can be that practices become more effective. These sorts of practice may improve threat modeling overall, leveraging their relative light by helping more
people threat model or threat model more often. The next section builds a conceptual framework for “fast, cheap, and good enough” threat modeling practices.


\subsection{Threat modeling: Responding to diverse needs}
\label{sec:orgb17836d}

Given the diverse goals of software and the diverse contexts in which software tools are
used, comprehensive “security checklists” become untenably tedious. As a result, threat
modeling seeks to imagine possible threats through speculative processes. For example,
STRIDE was originally proposed as a model of threats to products (STRIDE is an acrostic of
Spoofing, Tampering, Repudiation, Information Disclosure, Denial of Service and Elevation
of Privilege) (\cite{kohnfelder1999threats}). Engineers would consider those threats to a product as a whole. A derivative, STRIDE per element, examines each component of a system for relevant STRIDE threats (\cite{shostack2008security}). Clearly, the later use is both more time consuming than the former, and less likely to miss issues. Systems such as STRIDE are intended to help security workers search for attacks by providing prompts. (\cite{Shostack2014})
Board and card games sometimes assist in the speculative process of threat modeling (\cite{Denning2013} \cite{8194898} ). For example, the game \textit{Elevation of Privilege} stemmed from Microsoft's development of the STRIDE methodology (\cite{Shostack2014}). Even the US Central Intelligence Agency (CIA) developed a (now-declassified) game for helping developers generate threat models (\cite{Masnick2018}).

There are some common threads in critiques of existing methods:
\label{section:critiques}
\begin{enumerate}
\item \textbf{They are difficult to learn, especially for those who are not security specialists} (\cite{Weir2018,Denning2013}).
Developers who are not specifically trained in computer security
might write more secure software if they could better participate in threat identification. However, existing practices are generally aimed at security specialists
working at technology producers.
\item \textbf{They require significant time, even for specialists, to carry out.}
Threat modeling
methods require significant, devoted time. Even games like Elevation of Privilege
require at least 30 minutes; more detailed threat modeling practices can take hours,
days or weeks to complete.
\item \textbf{They may systematically fail to identify particular types of threats},
particularly those that arise from social factors or relational characteristics such as gender, race, age, or disability; or those whose impacts fall on end users or those who are invisibly impacted by it. (For example, the use of facial recognition on city streets.)
(\cite{Pierce2018,Freed2018,Nissenbaum2005,Hong2004,Friedman2012,Suchman2017,Coles-Kemp2009}).
\end{enumerate}

Given the variety of software systems and delivery methods, it seems unreasonable to
assume that there will be a single approach to modeling threats, any more than there will be a Unified Development Process.
Many threat modeling approaches have been proposed and are coalescing around an
organizing set of questions: ‘what are we working on,’ ’what can go wrong’, ‘what are we going to do’ and ’did we do a good job.’ (A set of industrial practitioners converged on these in a “Threat Modeling Manifesto” in 2020\footnote{Available at \url{threatmodelingmanifesto.org}}.)

\section{Emerging techniques serve a broader audience}

Many organizations are not doing security work at all. They are not threat modeling, using static analysis, or fuzzing. “Nothing” is good enough for them. In that context, fast and cheap methods may literally be better than that `nothing.' A more nuanced view is the adoption and use of security improvements is a complex tradeoff space, and we’re not good at evaluating these lightweight tools (\cite{hollnagel2006resilience}). This work provides three short case studies of lightweight approaches to threat modeling
presented elsewhere in the literature. These are intended to show approaches to security in product development that are lighter than commonly cited approaches. We can compare these to `STRIDE per element,’ a more structured approach. Similarly, for static analysis we compare it to sound approaches. (We include an example of a static analysis tool to show that present work is not restricted to threat modeling.)

\subsection{Security Fictions}

\subsubsection{What is Security Fictions?}
Security Fictions is an “improvisational role-playing game” that takes place between an
interlocutor (in the case of the study, the researcher) and an engineer. The interlocutor
poses as a system’s enemy: someone who wants to accomplish a particular goal using the
system. Examples from the study include (1) impersonating another user on the platform,
(2) finding a user’s geographic location, and (3) identifying all the users who have
particular political goals. The engineer then role-plays as someone trying to help the
enemy—an “insider threat”—and brainstorms with the interlocutor about how best to
achieve the nefarious goal. (\cite{Merrill2020})

\subsubsection{Why was Security Fictions designed this way?}

Security Fictions was designed to respond to critiques (1) and (2) in section~\ref{section:critiques}.
To make the game easier for non-specialists to learn, it looked toward another methodological tradition for surfacing potential future issues: speculative design. (In building this analogy, the authors observe “that threat identification is an intrinsically speculative practice: it
requires imagining possible futures.”) To synthesize these two research traditions, Security Fictions draws the threat modeling practice of redteaming together with Elsden et al.’s Speculative Enactments, which engage research participants in improvisational play with real-world consequences (\cite{Elsden2017}). 
From this role-playing tradition, the practice was designed to be quick to learn, and better at surfacing sociotechnical threats; from this redteaming tradition, it was designed to identify actionable and specific security issues.

\subsubsection{How well does Security Fictions work?}
In an exploratory deployment of the game among software developers (some of whom did
not specialize in security), developers enjoyed playing the game, and used it to come up with realistic threats. Some of the threats that non-security specialists devised were not strictly technical (for example, a UX designer discovered a social attack that relied on a misleading feature in a user interface).

Security Fictions provokes useful insights through two related mechanisms: gamification
and permission. The gamification is the explicit framing as a role playing game. This is accompanied by an invitation to “set ethical concerns aside,” giving social permission to discuss unwanted behaviors. This allows those with less explicit security knowledge to explore possible security flaws, even if they lack terminology, structure or other explicit security knowledge.

\subsubsection{Analysis}
Security Fictions are inexpensive to run (from 15-60 minutes per session) once a set of scenarios has been created or adopted. The open-ended nature of the framework allows
the use of fewer or more questions in a given session. There is no particular expertise
required to create scenarios, but a degree of either storytelling or dungeon-mastering
experience may be helpful in effective execution, and more research may reveal criteria for
good prompts.

\subsection{Elevation of Privilege}
\subsubsection{What is EoP?}
Elevation of Privilege is a threat modeling approach embedded into a card game. The deck
consists of 6 suits, based on the STRIDE model of threats. Within each suit there are cards 2
through Ace. Each card has a hint, for example the 8 of Information Disclosure says: ‘An
attacker can access information through a search indexer, logger, or other such
mechanism.’ The deck is used with a system diagram to ‘draw developers into threat
modeling,’ using familiar card game mechanics.

\subsubsection{Why was EoP designed this way?}
The game was designed to be an enticing, supportive and non-threatening way to draw
people into threat modeling, to bring in people with different skill sets and knowledge. I
wrote ‘If it were possible to have developers do basic threat modeling, then experts could
be used more effectively to find the really unusual problems with a design,’ and ‘Those who
are new to threat modeling or those who threat model occasionally require a more procedural approach, and procedural approaches are generally at odds with creativity. Elevation of Privilege was created in this constraint space in part to expose non-security experts to the enjoyment that security experts bring to threat modeling.’

\subsubsection{How well does EoP work?} The game has been available for a dozen years, and I reported on a variety of anecdotes to
its success ranging from six-year-olds to companies that report having weekly sessions (\cite{Shostack2014}.
There are at least three substantiative variants, including OWASP Cornucopia, F-Secure’s
Elevation of Privacy (with 4 new suits) and LogMeIn has added a privacy suit, making
STRIPED.
Elevation of Privilege enables contributions from people who lack security knowledge via
the hints on the cards, and requires it through the turn-taking game structure. The hints
are augmented by game-table conversation and banter, offering a chance for participants to
ask questions about their hints in a lower-pressure environment.
\subsubsection{Analysis} 
When Elevation of Privilege was created, it was intended to be the simplest, lowest cost
structured introduction to threat modeling possible. Since then, faster and simpler
approaches have appeared, putting the game at the more expensive end of fast, cheap approaches, taking about an hour.

\subsection{Semgrep}
We include discussion of semgrep because it may be to static analysis what Elevation of
Privilege is to threat modeling: a lighter, more accessible approach, and it may be helpful to
consider fast, cheap, good approaches beyond threat modeling.

\subsubsection{What it is?} Semgrep is an open source, lightweight static analysis tool that's designed to be fast, easy to deploy and easy to program. 

\subsubsection{Why was it designed this way?} ``Semgrep is designed for the security engineer or developer who wants to automate code review.'' and ``We think modern static analysis tools should run on every keystroke in the editor, without needing network access.'' (Both \url{https://semgrep.dev/docs/faq/}) 

These properties of easy to deploy, fast, and easy to write rules for seem intuitively useful, but they are less prioritized than soundness and completeness. Semgrep explicitly rejects the need to be able to handle an abstract syntax tree, stating in their FAQ  ``if you can write code, you can write a Semgrep rule — no program analysis PhD required!''.

\subsubsection{How well does it work?} Semgrep has a `trophy case' on their website with 8 CVEs, and a set of useful contributions, but its use seems to be largely internal to enterprises, which makes assessment more challenging. Further, the team behind it seems to have a frame that value is evidenced by continued use, rather than publications. As such, the rate of rule publication may be helpful evidence of engagement. As of April, 2021, Semgrep's core was being released approximately weekly, in addition to roughly 25 new rules contributed each week. This indicates that at least a few people are motivated to continue participating.

\subsubsection{Relevant reflections} 
Semgrep explicitly focuses on being inexpensive to learn and use, without specifying what those mean in practice. They also focus on being accessible to those without a PhD.

The relationship between applied and academic static analysis work was the subject of reflection by the Coverity team. Their paper on the subject discusses many of the challenges that semgrep has taken on, including troubles in deploying their tool, speed, and managing multiple languages (\cite{billionlineslater}).

\section{A Flaw Space and A Flaw Oracle}

\begin{figure*}
    \centering
    \includegraphics[width=\textwidth]{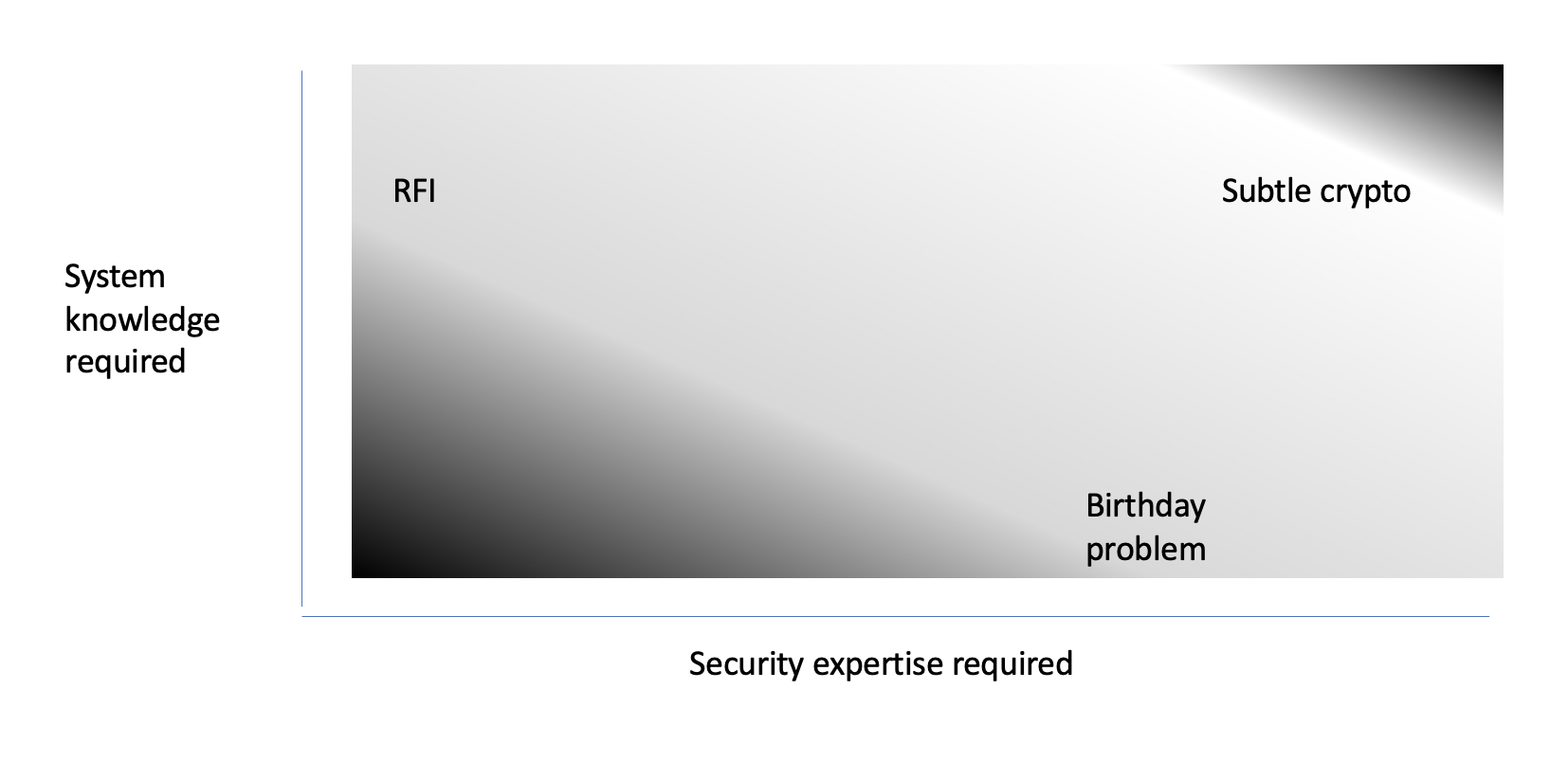}
    \caption{Figure 1}
    \label{fig:theworld}
\end{figure*}

The distribution of flaws by the system knowledge and security expertise required to find them (Darker areas have more flaws). Issues in the lower left are what we describe as \textit{light flaws}. Lightweight threat modeling practices can help find them more effectively than the widely-recommended practices common today.

Flaws require different levels of system knowledge, and different levels of security expertise, to discover.
Some flaws, which we will herein refer to as \textit{light flaws}, require a less system knowledge and/or less security expertise to uncover (Figure \ref{fig:theworld}). We argue that the methods we describe above work because they help to catch light flaws, and do so with a minimum of time and training on the part of developers.

Existing threat modeling practices, we argue, aim to capture the entire flaw space. In doing so, they undervalue the time, energy, and organizational costs required to do so. By providing lighter practices by which non-specialists can capture and solve light flaws, security experts would be relatively less burdened, able to focus only on a subset of the flaw space (represented by the upper-right in Figure \ref{fig:theworld}).
The remainder of this section uses a thought experiment to expound on our reasoning.

Imagine a perfect examiner who instantly and reliably finds flaw in a system, given the correct inputs. This Flaw Oracle takes as input first, knowledge of the system and second, knowledge of security (This is a thought experiment, we don't propose such an examiner exists, nor do we claim such knowledge can be codified). We are focused on what flaws will be found, without specifying an underlying distribution, or requiring knowledge of the actual distribution.  We can also imagine that an examiner requires intensive security knowledge. Informally, some theorem provers might be analogous to such an examiner. (The term `flaw' is often used for the design subset of security issues, complementing bugs or vulnerabilities. We use it to encompass both, as \emph{security oracle} is too broad. We can also consider a Laggard Flaw Oracle (LFO), who finds flaws and emits information at a later date. 

The knowledge needed by the examiner can be seen as analogous to the participants in system engineering. The software engineers may have a great deal of knowledge of the system, and less knowledge of security. Security engineers, especially consultants, will bring knowledge of security, and less knowledge of the system.
Knowledge is expensive. Each step of gathering it, checking it, encoding it into human- or machine-readable forms, checking the encoding didn't introduce problems takes effort. Similarly, keeping such information up-to-date as systems evolve requires effort.

We can visualize the distribution of flaws in systems, and simple visualizations such as Figure \ref{fig:theworld} can capture intuitions. Another  possible distribution (not shown) is that flaws are distributed randomly and evenly.

Thus systems that require fewer expensive inputs can have value, if there are many issues that an examiner can find with system knowledge and a modest amount of security knowledge. For example, once you know that file APIs will often accept network paths, someone with knowledge of where a system opens files can easily check for remote file inclusion (RFI) bugs. The birthday problem is easily taught, but seeing how it applies to a system can be harder, and subtle cryptographic bugs may require both security and system knowledge.

This account of the flaw space explains \textit{why} the lightweight threat modeling practices we describe above may have worked.
Further, it provides a theoretical backdrop that motivates further work on developing and evaluating lightweight threat modeling practices.
Readers may be concerned that this work will do more harm than good. An analog of Gresham's Law of ``Bad money drives out good'' may be relevant: bad security may drive out the good. We disagree. The availability of inexpensive clothes does not stop people from paying for designer clothes; each delivers a different value. 

\section{Evaluating Practices}

Having put forth the idea that light flaws exist, we ask: how best can we judge practices that
aim to catch these flaws? We propose three, primary dimensions: cost,
expertise requirements, and quality of output. These are important dimensions which
make up a multidimensional design space; we expect there are other useful criteria beyond
these. These criteria exclude the impact of the output, which has surprised some early readers. The
usefulness of a technique is in a context, and the tradeoffs which that context demands are
independent of the tool (We do not judge a hammer useless because we have a lot of hex
nuts).

\subsection{Cost} 
Cost include money, time, and organizational energy. Time includes training, preparation,
and execution. Some systems require only minutes of training or setup. For example,
Security Fictions takes a few minutes to explain and set the stage. Elevation of Privilege
usually involves a few minutes of stage-setting, 5-15 minutes to draw a system diagram,
and a round of test play for people to ‘get the hang of it.’ This is in contrast to a half day or
even days of training in deeper threat modeling approaches. Some systems, like Security
Fictions, requires some prep work in crafting the scenarios. (It is reasonable to think that a
compendium of scenarios could be created, but none has.) Organizational costs include
coordination overhead associated with communicating flaws across functional teams,
prioritizing issues to fix, and so on. The more that lightweight processes reduce
coordination overhead (e.g., by empowering UX designers to solve problem on their own),
the lower their cost. The LFO adds cost by delay. It is well understood that fixing issues is more expensive after code is written, and more expensive yet when the code has dependencies which need to be at least checked and possibly fixed, or when the code is already deployed possibly at many sites, each of which must deploy the fixes. Thus, the slower the LFO, the more cost it allows to creep into the system.
\subsection{Expertise requirement}
Figure \ref{fig:theworld} shows a model where discovering flaws require different levels of system-specific
and security-specific expertise. Effective lightweight methods can minimize reliance on
security-specific expertise, compensating by relying instead on system-specific expertise.
In some empirical work (\cite{Merrill2019}), a UX designer used misleading unicode symbols to perform an
impersonation attack, effectively using their systems knowledge to compensate for a lack of
security-specific knowledge. The ability of various practices to facilitate this tradeoff is
critical for catching light flaws.
Characterizing and defining the exact security knowledge required to make good use of a
tool is beyond the scope of this paper, but we’ll note briefly that we have never seen a
characterization of skill except “this tool requires a PhD in static analysis.”

\subsection{Quality}
The most challenging but most crucial dimension to measure is the quality of practices' output. Students often seek to find ``good threats.'' What makes a ``good threat''? There is an argument that a ``good threat'' discovered by a threat modeling approach is one that the organization cares about enough to fix. That is, the time spent to fix it is evidence of a useful output. But there can be reasons to not fix a worrisome problem. The organization may be fixing other more pressing issues, there may be issues of application compatibility or usability of a fix. In other words, the perceived relevance of flaws is contextual. The exact same issue, discovered at (for example) Amazon, Microsoft, Bank of America or a local community bank might be treated differently at each. Bug bounties offered by each are scoped differently, rewarding different flaw types.

There may be work in assessing quality from the perspective of the users. For example, in (\cite{slupska2021participatory}) an argument is put forth for threat modeling techniques that are easier for those other than cybersecurity professionals to use. There are important dimensions of quality, such as avoiding creepy advertisements or stalking, that traditional information-security centered approaches may not prioritize.
Future work should delve into these questions as they apply to the flaw space described here.

\section{Future work}
We hope to create a framework in which future approaches to threat modeling can be understood, evaluated and compared. We hope to enable work in the area of fast and cheap tools to help people analyze security, and to enable rigorous and thoughtful discussion of those tools. Ease of use, speed, scope, suitability for non-technical staff at organizations building systems, suitability for use by the public, advocates, activists, and other factors are all important.
By defining the nature of tradeoffs that threat modeling tools make, we hope to enable exploration of new tools that more strategically and purposefully approach different portions of the flaw space we describe in Figure \ref{fig:theworld}.

While this work is highly preliminary, it raises one critical point of
reflection for future research: our observation that a flaw space, distributed
across system and security expertise, might exist begs a cultural shift in the
way that we, as security professionals, evaluate security processes and
outcomes.

How do we judge the quality of processes and outcomes? Analyses, typically
received from academia, focus on the soundness of analysis, formality of proofs,
precision and recall of algorithms, and so on. But perhaps quality in security
is an emergent characteristic of an organization. Banks and falafel stops have
not only different security needs, but different ways of evaluating what kinds
of security make sense for them to approach, and different competing constraints
for their time and attention.

Sensitivity to quality as an emergent and situated characteristic of
organizations allows us to judge threat modeling practices better. It centers
broad outcomes, and diverse notions of what it means to be secure.

The cultural shift here, for us, is to make less formal and less analytical
processes acceptable by the cultural standards of security research 
(The author has seen a paper rejected with the comment ``This is great, and it needs equations'').
Moving
toward these more situated notions will help us understand the work that, for
example, a local falafel shop does to protect themselves as relevant and
important, assessing the quality of their output that isn't overly tied to
narrow analytical frameworks. I return, this will help us build practices that
make security more relevant and actionable to the people who need it.

Our call to action: security needs better ways of meeting people where they are,
understanding what they need. An analogy to economics helps illustrate the
conceptual shift required here. In economics, a broad and useful assumption is
that people are rational; that they have a reason to do the things that they do.
When economists observe people doing something seemingly irrational (for
example, not saving as much as they ``should''), economists look for the
rationality that underlies that decision (for example, perhaps people don't save
because they believe they'll die having never spent their money; this makes
saving contextually irrational).

As security researchers, we can do something similar. What are people
\textit{doing}, and what is the rationality that underlies those actions? When
people ignore our security advice, perhaps doing so is rational for them
because, for example, it costs them too much in time or energy relative to their
other goals (see our design dimensions above and \cite{herley2009so}). How can we best use those
observations to design more relevant and actionable security practices? 

\section{Acknowledgements}
This article comes from deep conversation with Nick Merrill about why these lightweight approaches are meaningful. The author appreciates thoughtful comments from Mary Ellen Zurko, which improved this draft.









\bibliographystyle{unsrtnat}
\bibliography{references}  






\end{document}